\documentclass[amsfonts,12pt]{article}

\usepackage{amsmath, amssymb, amsfonts, amsthm, fouriernc}
\title{}\date{}

\title{Friedmann's Equations in All Dimensions\\ and  Chebyshev's
Theorem}

\author{Shouxin Chen\footnote{Email address: chensx@henu.edu.cn}\\Institute of Contemporary Mathematics\\School of Mathematics and Statistics\\Henan University\\
Kaifeng, Henan 475004, PR China\\\\
Gary W. Gibbons\footnote{Email address: gwg1@damtp.cam.ac.uk}\\
D. A. M. T. P.\\
University of Cambridge\\
Cambridge CB3 0WA, U. K.\\\\Yijun Li\footnote{Email address: liyijun@henu.edu.cn}\\School of Mathematics and Statistics\\Henan University\\
Kaifeng, Henan 475004, PR China\\\\Yisong Yang\footnote{Email address: yisongyang@nyu.edu}\\Department of Mathematics\\Polytechnic School, New York University\\Brooklyn, New York 11201, U. S. A\\ \&\\NYU-ECNU 
Institute of Mathematical Sciences\\New York University - Shanghai\\3663 North Zhongshan Road, Shanghai 200062, PR China}
\newcommand{\bfR}{{\Bbb R}}

\def\XXint#1#2#3{{\setbox0=\hbox{$#1{#2#3}{\int}$}
 \vcenter{\hbox{$#2#3$}}\kern-.5\wd0}}

\newtheorem{oldtheorem}{Theorem}[section]
\newtheorem{oldassertion}[oldtheorem]{Assertion}
\newtheorem{oldproposition}[oldtheorem]{Proposition}
\newtheorem{oldremark}[oldtheorem]{Remark}
\newtheorem{oldlemma}[oldtheorem]{Lemma}
\newtheorem{olddefinition}[oldtheorem]{Definition}
\newtheorem{oldclaim}[oldtheorem]{Claim}
\newtheorem{oldcorollary}[oldtheorem]{Corollary}

\newcommand{\dd}{\mbox{\footnotesize    d}}
\newcommand{\ee}{\end{equation}}
\newcommand{\be}{\begin{equation}}\newcommand{\bea}{\begin{eqnarray}}
\newcommand{\eea}{\end{eqnarray}}
\newcommand{\e}{\mbox{\footnotesize    e}}
\newcommand{\Om}{\Omega}

\newcommand{\nn}{\nonumber}

\begin{document}
\maketitle

\newpage

\begin{abstract}
This short but systematic work demonstrates a link between   Chebyshev's  theorem and the explicit integration
in cosmological time $t$ and conformal time $\eta$ of the Friedmann equations
in all dimensions and with an arbitrary cosmological constant $\Lambda$.
More precisely, it is shown that for spatially  flat universes  an explicit
integration
in $t$ may always be carried out, and that,
in the non-flat situation and when $\Lambda$ is zero
and
the ratio $w$ of the pressure and energy density in the barotropic equation of state of the perfect-fluid universe is rational,  an explicit integration may be carried out if and only if the dimension $n$ of space and $w$ obey some specific relations among an infinite family. The situation for explicit integration in $\eta$ is complementary to
that in $t$. More precisely, it is shown in the flat-universe case
with $\Lambda\neq0$ that an explicit integration in $\eta$  can
be carried out if and only if $w$ and $n$ obey similar
relations among a well-defined  family which we specify,
and that, when $\Lambda=0$, an explicit
 integration can always be carried out whether the
space is flat, closed, or open. We also  show that
our method may be used to study more realistic
cosmological situations when the equation of state is nonlinear.
\medskip

{\bf Keywords:} Astrophysical fluid dynamics, cosmology with extra dimensions,
alternatives to inflation,
initial conditions and eternal universe,
cosmological applications of theories with extra dimensions,
string theory and cosmology.

\medskip

{ PACS numbers:} 04.20.Jb, 98.80.Jk

\end{abstract}

\maketitle

\section{Introduction}
\setcounter{equation}{0}

The Friedmann equations play a fundamental role in modern cosmology, in particular for studies  of  inflationary universes.
One of the main tools for simplification of these equations, in order to obtain an exact integration, is through the use of conformal-time reparametrization
\cite{Barrow,Gibbons,SK,F,L,UL,Linder,CDS,van,W}, although qualitative investigations are often more  conveniently  carried out using
 cosmological time \cite{Harrison,ST,WKS,FS0,CS,sc1}.
Particularly noteworthy is the classification scheme of  Harrison
\cite{Harrison} which contains many explicit formulas
for the scale factor as a function of cosmic time. However
the range of possible  equations of state treated is rather limited,
attention is restricted to the three-dimensional case and
the paper is not directly concerned with
finding all cases which can be integrated
in finite terms.
Here we show that, in view of  Chebyshev's  theorem,
the answer to this question: whether the Friedmann equations may be
integrated  explicitly in cosmological time
becomes
immediately transparent,
at least for two important situations in all dimensions:
spatially flat universes   with an arbitrary cosmological constant
and spatially curved  universes  with vanishing cosmological constant.
 More precisely, we show that, in the former situation,
an explicit  integration \footnote{By an explicit integration
we mean what is often referred to as an integration in finite
terms or an analytic integration or an
integration in terms of elementary functions.
From now on we shall drop the adjective
explicit and integration should be taken to mean in finite terms.} may be carried out
when the ratio $w$ of the pressure and energy density in the barotropic equation of state
of the perfect-fluid universe is arbitrary, and in the latter situation and when $w$ is rational,
an integration may be carried out if and only if $w$
assumes specific values among an infinite sequence.

Even though nowadays, numerical solutions of ordinary differential equations
are readily available on laptops and other
electronic devices, it is still the case
 that, for a rapid overview of the
properties of the  solutions and their
detailed behavior, it is  often more
efficient to examine  explicit formulas
when they are available. Hence this paper.

Here is an outline of the rest of the paper. In Section 2 we recall  Friedmann's equations and Chebyshev's theorem and discuss the applicability of the latter to the former. In
Section 3 we show that, in view of Chebyshev's theorem, an explicit integration of Friedmann's equations
in cosmological time may always be carried out in all spatial dimensions and for any value of the
ratio $w$ of the pressure and energy density in the equation of state of the perfect-fluid universe. In Section 4 we tackle the non-flat situations assuming the cosmological constant vanishes
and we identify an infinite family of values of $w$ that permit an explicit integration in cosmological time. In Section 5 we consider the  problem in conformal time and determine all the cases when
an explicit integration can be carried out. This situation is also shown to 
complement that in cosmological time.

\section{Friedmann's equations and Chebyshev's theorem}
\setcounter{equation}{0}

Consider an $(n+1)$-dimensional  homogeneous and isotropic
Lorentzian
spacetime with the metric
\be \label{1}
\dd s^2=g_{\mu\nu}\dd x^\mu\dd x^\nu=-\dd t^2+a^2(t)g_{ij} \dd x^i\dd x^j,\quad i,j=1,\dots,n,
\ee
where $t$ is the cosmological (or cosmic) time and $g_{ij}$ is the  metric of an $n$-dimensional Riemannian manifold
$M$ of constant scalar curvature characterized by an indicator, $k=-1,0,1$,
so that $M$ is an $n$-hyperboloid, the flat space $\bfR^n$, or an $n$-sphere, with the
respective metric
\be \label{2}
g_{ij}\dd x^i\dd x^j=\frac1{1-kr^2}\,\dd r^2+r^2\,\dd\Om^2_{n-1},
\ee
where $r>0$ is the radial variable and $\dd\Om_{n-1}^2$ denotes the canonical metric of the unit
sphere $S^{n-1}$ in $\bfR^n$. Inserting the metric (\ref{1})--(\ref{2}) into the Einstein equations
\be
G_{\mu\nu}+\Lambda g_{\mu\nu}=8\pi G_n T_{\mu\nu},
\ee
where $G_{\mu\nu}$ is the Einstein tensor, $G_n$ the universal gravitational constant in $n$ dimensions, and $\Lambda$ the cosmological constant, the speed of light is set to unity, and $T_{\mu\nu}$ is the
energy-momentum tensor of an ideal cosmological fluid given by
\be \label{4}
T_{\mu}^{\nu}=\mbox{diag}\{-\rho_m,p_m,\dots,p_m\},
\ee
with $\rho_m$ and $p_m$ the $t$-dependent matter energy density and pressure,
we arrive
at the Friedmann equations
 \bea
H^2&=&\frac{16\pi G_n}{n(n-1)}\rho-\frac k{a^2},\label{5}\\
\dot{H}&=&-\frac{8\pi G_n}{n-1}(\rho+p)+\frac k{a^2},\label{6}
\eea
in which
\be \label{7}
H=\frac{\dot{a}}a,
\ee
denotes the usual Hubble `constant', $\dot{f}=\frac{\small{\dd} f}{\small{\dd} t}$, and $\rho,p$ are the effective energy
density and pressure related to $\rho_m,p_m$ through
\be \label{8}
\rho=\rho_m+\frac{\Lambda}{8\pi G_n},\quad p=p_m-\frac{\Lambda}{8\pi G_n}.
\ee
On the other hand, recall that, with (\ref{1}) and (\ref{4}) and (\ref{8}), the energy-conservation law, $\nabla_\nu T^{\mu\nu}=0$, takes the form
\be \label{9}
\dot{\rho}_m+n(\rho_m+p_m)H=0.
\ee
It is readily seen that (\ref{5}) and (\ref{9}) imply (\ref{6}). In other words, the full cosmological
governing equations consist of (\ref{5}) and (\ref{9}) only \footnote{In the cosmological literature
(\ref{5}) is often referred to as \emph{the} Friedmann equation
and (\ref{6}) somewhat anachronistically  as the Raychaudhuri equation.
In this paper by the  Friedmann equations we mean
the pair (\ref{5}), (\ref{9}), supplemented with an equation of state, $p_m=
p_m(\rho_m)$.}   .

To proceed further we assume a  so-called
barotropic equation of state \footnote{Later we will relax this assumption
slightly by considering a mixture of two barotropic fluids with different
values of $w$, but set the cosmological constant to zero} ,
\be \label{10}
p_m=w \rho_m,
\ee
where $w$ is a constant so that $w=0$ leads to a vanishing pressure, $p_m=0$, corresponding to
the dust model,
and $w=\frac1n$, the radiation-dominated model.

Inserting (\ref{10}) into (\ref{9}), we have
\be\label{11}
\dot{\rho}_m+n(1+w)\rho_m \frac{\dot{a}}a=0,
\ee
which can be integrated to yield
\be\label{12}
\rho_m=\rho_0 a^{-n(1+w)},
\ee
where $\rho_0>0$ is an integration constant \cite{KL}. Using (\ref{12}) in (\ref{8}), we arrive at
the relation \cite{NOO,KL}
\be\label{13}
\rho=\rho_0 a^{-n(1+w)}+\frac{\Lambda}{8\pi G_n}.
\ee
From (\ref{5}) and (\ref{13}), we get the following equation of motion for the scale factor $a$:
\be \label{14}
\dot{a}^2=\frac{16\pi G_n\rho_0}{n(n-1)}a^{-n(1+w)+2}+\frac{2\Lambda}{n(n-1)} a^2-k.
\ee

To integrate (\ref{14}), we recall Chebyshev's  theorem
\cite{C0,C1}: For rational numbers
$p,q,r$ ($r\neq0$) and nonzero real numbers $\alpha,\beta$, the integral
\be\label{IaI}
I=\int x^p(\alpha+\beta x^r)^q\,\dd x
\ee
 is elementary
if and only if at least one of the quantities
\be\label{cd}
\frac{p+1}r,\quad q,\quad \frac{p+1}r+q,
\ee
is an integer.
In fact, the integral (\ref{IaI}) may be rewritten as \cite{planetmath}
\bea
I&=& \frac{1}{r} \alpha ^{\frac{p+1}{r}+q} \beta ^{-\frac{p+1}{r }}
B_y \left(\frac{1+p}{r}, q-1\right)\nn\\
&=&
\frac{1}{p+1} \alpha ^{\frac{p+1}{r}+q}
\beta ^{-\frac{p+1}{r} }  y^{\frac{1+p}{r}}
F\left(\frac{p+1}{r},2-q , \frac{1+p+r}{r};y\right) \,,
\eea
where $ y= \frac{\beta }{\alpha} x^r $, and  $B_y \left(\frac{1+p}{r}, q-1\right)$
and $F\left(\frac{p+1}{r},2-q , \frac{1+p+r}{r};y\right)$
are the incomplete beta function and hyper-geometric function, respectively.
The result then follows from the classification of elementary cases of the
hyper-geometric function.

 Consequently, when $k=0$ or $\Lambda=0$, and $w$ is rational, the Chebyshev theorem enables us to know that,
for exactly
what values of $n$ and $w$, the equation (\ref{14}) may be integrated.

\section{Spatially flat case}
\setcounter{equation}{0}

We first focus on the spatially flat situation $k=0$, which is known to be
most relevant  for cosmology \cite{Planck,P2,P3,nature,science,Copeland},
and rewrite  equation (\ref{14}) as
\be \label{15}
\dot{a}=\pm\sqrt{c_0 a^{-n(1+w)+2}+\Lambda_0 a^2},\quad c_0=\frac{16\pi G_n\rho_0}{n(n-1)},\quad\Lambda_0=\frac{2\Lambda}{n(n-1)}.
\ee
(This equation, when $n=3$, arises in the study of holographic cosmology \cite{KL} [see particularly
the words under (15) in \cite{KL}], which initiated the  present 
systematic approach.) Formerly, (\ref{15}) reads
\be \label{15a}
\pm\int a^{-1}\left(c_0 a^{-n(1+w)}+\Lambda_0 \right)^{-\frac12}\dd a=t+C.
\ee
It is clear that the integral on the left-hand side of (\ref{15a}) satisfies the integrability condition stated in the Chebyshev theorem for any $n$ and any rational $w$.

We have just seen that (\ref{15}) can be integrated directly for any rational $w$. This indicates
that (\ref{15}) might be integrated for any real $w$  as well,
not necessarily rational.
Indeed, we show that this is true. To do so we apply $a>0$ and get from (\ref{15}) the equation
\be \label{16}
\frac{\dd}{\dd t}\ln a=\pm\sqrt{c_0 a^{-n(1+w)}+\Lambda_0},
\ee
or equivalently,
\be\label{17}
\dot{u}=\pm\sqrt{c_0\e^{-n(1+w)u}+\Lambda_0},\quad u=\ln a.
\ee
Set
\be\label{18}
\sqrt{c_0\e^{-n(1+w)u}+\Lambda_0}=v.
\ee
Then
\be\label{19}
u=\frac{\ln c_0}{n(1+w)}-\frac 1{n(1+w)} \ln(v^2-\Lambda_0).
\ee
Inserting (\ref{19}) into (\ref{17}), we find
\be\label{20}
\dot{v}=\mp\frac12 n(1+w)(v^2-\Lambda_0),
\ee
whose integration gives rise to the expressions
\be \label{21}
v(t)=\left\{\begin{array}{rl} &v_0\left(1\pm \frac12 n(1+w)v_0t\right)^{-1},\quad \Lambda_0=0;\\
&\\
& \sqrt{\Lambda_0}(1+C_0\e^{\mp n(1+w)\sqrt{\Lambda_0} t})(1-C_0\e^{\mp n(1+w)\sqrt{\Lambda_0} t})^{-1},\\& C_0=(v_0-\sqrt{\Lambda_0})(v_0+\sqrt{\Lambda_0})^{-1},\quad \Lambda_0>0;\\
&\\
&\sqrt{-\Lambda_0}\tan\left(\mp\frac12 n(1+w)\sqrt{-\Lambda_0} t +\arctan\frac{v_0}{\sqrt{-\Lambda_0}}\right),\quad
\Lambda_0<0,
\end{array}
\right.
\ee
where $v_0=v(0)$. Hence, in terms of $v$, we obtain the time-dependence of the scale factor $a$:
\be \label{22}
a^{n(1+w)}(t)=\frac{8\pi G_n \rho_0}{\frac12 n(n-1)v^2(t)-\Lambda}.
\ee

We now demonstrate how to use the analytic solutions we have
obtained to study cosmology. For definiteness, we assume
\be
w>-1
\ee
in the equation of state in our subsequent discussion.
We are particularly interested in solutions satisfying
\be\label{a0}
a(0)=0,
\ee
which represents a universe evolving from a single point singularity.


In view of (\ref{21}), (\ref{22}), and (\ref{a0}), we get
\bea \label{x1}
a^{n(1+w)}(t)=\left\{\begin{array}{rll}&4\pi G_n\rho_0\left(\frac n{n-1}\right)(1+w)^2 t^2,&\Lambda=0;\\
&\frac{8\pi G_n\rho_0}{\Lambda}\sinh^2\left(\sqrt{\frac{n\Lambda}{2(n-1)}}(1+w) t\right),&\Lambda>0.\end{array}\right.
\eea
Both cases lead to an expanding universe.

We now consider the case when $\Lambda<0$ separately and rewrite (\ref{22}) as
\be \label{23}
a^{n(1+w)}(t)=\frac{8\pi G_n \rho_0}{(-\Lambda)}\cos^2\left(\sqrt{\frac{n(-\Lambda)}{2(n-1)} }(1+w)t \mp\arctan\sqrt{\frac{n(n-1)}{-2\Lambda}}\, v_0\right).
\ee
If we require $a(0)=0$, then (\ref{23}) leads to the conclusion
\be \label{24}
a^{n(1+w)}(t)=\frac{8\pi G_n \rho_0}{(-\Lambda)}\sin^2\sqrt{\frac{n(-\Lambda)}{2(n-1)} }(1+w)t,
\ee
which gives rise to a periodic universe.

The above solutions cover those obtained in the case $n=3$ presented in \cite{SK,AA}.


\medskip

{\bf Application to  cosmic jerk}
\medskip

Needlessly to say, with the formulas (\ref{x1}) and (\ref{24}), various quantities of cosmological
interest may be calculated exactly. For example, the quantity
\be
\label{4.1}
Q=a^2\left(\frac{\dd a}{\dd t}\right)^{-3} \frac{\dd^3 a}{\dd t^3}
\ee
is listed \cite{G2} as one of four cosmological scalars known as jerk.  In order
to facilitate the computation, we may represent the scale factor as $a(t)=c f^s(t)$, where $c,s>0$ are constants, and insert it into (\ref{4.1}) to obtain
\be\label{4.2}
Q=\frac{(s-1)(s-2)}{s^2}+\frac{3(s-1)}{s^2}\frac{f\ddot{f}}{(\dot{f})^2}+\frac1{s^2}\frac{f^2{\dddot f}}{(\dot{f})^3}.
\ee
Thus, in view of (\ref{4.2}) and the expressions given in (\ref{x1}) and (\ref{24}), we may insert
\be\label{4.3a}
f(t)=\left\{\begin{array}{rll}&t,\quad & \Lambda=0,\\ &\sinh\sqrt{\frac{n\Lambda}{2(n-1)}}(1+w)t,\quad
&\Lambda>0,\\&\sin\sqrt{\frac{n(-\Lambda)}{2(n-1)}}(1+w)t,\quad&\Lambda<0,\end{array}\right.
\ee
into (\ref{4.2}) to obtain
\be\label{4.3}
Q=\left\{\begin{array}{rll}&\frac{(s-1)(s-2)}{s^2},\quad & \Lambda=0,\\ &\frac{(s-1)(s-2)}{s^2}+\frac{(3s-2)}{s^2}\tanh^2\sqrt{\frac{n\Lambda}{2(n-1)}}(1+w)t,\quad
&\Lambda>0,\\ &\frac{(s-1)(s-2)}{s^2}-\frac{(3s-2)}{s^2}\tan^2\sqrt{\frac{n(-\Lambda)}{2(n-1)}}(1+w)t,\quad&\Lambda<0,\end{array}\right.
\ee
where
\be
s=\frac2{n(1+w)}.
\ee

From (\ref{4.3}), we see clearly that the only $t$-independent jerk $Q$, uniformly valid for any cosmological constant $\Lambda$, is when $s=\frac23$, thus $Q\equiv1$, which leads to the condition
\be
n(1+w)=3. \label{f}
\ee
 This condition spells out a relation between the spatial dimension and the ratio of the pressure and energy density in the equation of state. In particular, when $n=3$, we must have
the dust universe, $w=0$.

 It is a striking fact that our present
universe (for which $n=3$ is well approximated as being  spatially flat and  with matter content
given by pressure free matter, $w=0$, and a positive cosmological constant
is completely characterised by the statement $Q=1$, i.e  that
``the cosmic jerk is unity''.  From this point of view the cosmological constant
is just one of the three integration constants of the third-order
ordinary differential equation obtained
by equating  the right-hand side of (\ref{4.1}) to one.
Since (\ref{f}) implies that 
\be
w= \frac{3}{n}-1 \,,
\ee 
we see that in dimensions higher than $3$ the pressure is negative
and decreases monotonically to a  cosmological constant  or dark energy equation
of state, i.e. $w=-1$, as     $n$ tends to infinity.

 \section{Zero cosmological constant case}
 \setcounter{equation}{0}

We now turn our attention to the situation $\Lambda=0$ in (\ref{14}) so that the equation reads
\be \label{32}
\dot{a}^2=\frac{16\pi G_n\rho_0}{n(n-1)}a^{-n(1+w)+2}-k.
\ee

In order to apply Chebyshev's  theorem, we now assume that $w$ is rational. Thus we see that the question whether (\ref{32}) may be integrated in cosmological time is equivalent to whether
\bea
I&=&\int a^{\frac12 n(1+w)-1}\left(-k a^{n(1+w)-2}+\sigma\right)^{-\frac12}\,\dd a\nn\\
&=&\frac2{n(1+w)}\int \left(-k u^{\gamma}+\sigma\right)^{-\frac12}\,\dd u,\quad u=a^{\frac12 n(1+w)},\quad\sigma=\frac{16\pi G_n\rho_0}{n(n-1)},\label{33}
\eea
is an elementary function of $u$, where
\be\label{34}
\gamma=2\left(1-\frac2{n(1+w)}\right).
\ee
By (\ref{34}), we see that (\ref{33}) is not elementary
unless $\frac1\gamma$ or $\frac{2-\gamma}{2\gamma}$ is an integer.\footnote{The case when $\gamma=0$ or $w=\frac{2-n}n$ is trivial since it renders $a(t)$ a linear function
through (\ref{32}).} That is, (\ref{32}) may only be
integrated directly in cosmological time when $w$ satisfies one of the following:
\bea
w&=&\frac{4N}{n(2N-1)}  -1,\quad N=0,\pm1,\pm2,\dots;\\
w&=&\frac2n+\frac1{nN}-1,\quad N=\pm1,\pm2,\dots.
\eea

In particular, in the special situations when $n=3$, we have
\be
w=-1,-\frac23,-\frac59,-\frac12,-\frac7{15},-\frac49,-\frac37,\dots,-\frac29,-\frac15,-\frac16,-\frac19,0,\frac13,
\ee
so that $-\frac13$ is the only limiting point.
Solutions of the Friedmann equations in
cosmological time corresponding to $w=0$ (dust model) and $w=\frac13$ (radiation model) are discussed in
\cite{SK}, among others.

As an illustration, we integrate (\ref{32}) for
\be
w=-\frac59,
\ee
($n=3$) so that (\ref{33}) becomes
\be
I=\frac32\int(-k u^{-1}+\sigma)^{-\frac12}\,\dd u,\quad u=a^{\frac23}.
\ee

When $k=1$ (closed universe), we may use the substitution $U=\sqrt{\sigma u-1}$
to carry out the integration, which gives us the explicit solution
\bea \label{41}
&&a^{\frac13}\sqrt{\sigma a^{\frac23}-1}-a_0^{\frac13}\sqrt{\sigma a_0^{\frac23}-1}
+\frac1{\sqrt{\sigma}}\ln\left(\frac{\sqrt{\sigma a^{\frac23}-1}+\sqrt{\sigma} a^{\frac13}}{\sqrt{\sigma a_0^{\frac23}-1}+\sqrt{\sigma} a_0^{\frac13}}\right)=\frac23\sigma t,\\
&&\quad t\geq0, \quad a_0=a(0),\nn
\eea
where $a_0$ satisfies the consistency condition $\sigma a_0^{\frac23}\geq1$ or
\be
a_0\geq\left(\frac3{8\pi G_3 \rho_0}\right)^{\frac32},
\ee
which spells out a minimum size of the universe in terms of $\rho_0$ whose initial energy density in view of (\ref{12}) is given by
\be
\rho_m(0)=\frac{64}9 \pi^2 G_3^2\rho^3_0.
\ee

When $k=-1$ (open universe), we may likewise use the substitution $U=\sqrt{\sigma u+1}$  to obtain the solution
\bea \label{44}
&&a^{\frac13}\sqrt{\sigma a^{\frac23}+1}-a_0^{\frac13}\sqrt{\sigma a_0^{\frac23}+1}
-\frac1{\sqrt{\sigma}}\ln\left(\frac{\sqrt{\sigma a^{\frac23}+1}+\sqrt{\sigma} a^{\frac13}}{\sqrt{\sigma a_0^{\frac23}+1}+\sqrt{\sigma} a_0^{\frac13}}\right)=\frac23\sigma t,\\
&&\quad t\geq0, \quad a_0=a(0),\nn
\eea
where no restriction is imposed on the initial value of the scale factor $a=a(t)$. In particular, if
we adopt the big bang scenario, we can set $a_0=0$ to write down the special solution
\be
a^{\frac13}\sqrt{\sigma a^{\frac23}+1}
-\frac1{\sqrt{\sigma}}\ln\left({\sqrt{\sigma a^{\frac23}+1}+\sqrt{\sigma} a^{\frac13}}\right)=\frac23\sigma t,\quad t\geq0.
\ee

Of course, the solutions (\ref{41}) and (\ref{44}) may  collectively and  explicitly be recast in the form of an elegant single formula:
\bea \label{44a}
&&a^{\frac13}\sqrt{\sigma a^{\frac23}-k}-a_0^{\frac13}\sqrt{\sigma a_0^{\frac23}-k}
+\frac{k}{\sqrt{\sigma}}\ln\left(\frac{\sqrt{\sigma a^{\frac23}-k}+\sqrt{\sigma} a^{\frac13}}{\sqrt{\sigma a_0^{\frac23}-k}+\sqrt{\sigma} a_0^{\frac13}}\right)=\frac23\sigma t,\\
&&\quad t\geq0, \quad a_0=a(0),\quad k=\pm1.\nn
\eea

We see that, in both closed and open situations, $k=\pm1$, respectively, the universe grows following a power law
of the type $a(t)=\mbox{\footnotesize    O}(t^{\frac32})$
for all large time so that
a greater Newton's constant or initial energy density gives rise to a greater growth rate.

\medskip

{\bf Application: dust and radiation}
\medskip

To see how we may apply  Chebyshev's  theorem to the study of
some realistic models in cosmology, we
consider $\Lambda=0,k=0$ and insert
\be
\rho=\frac{\rho_{\mbox{\footnotesize    m}}}{a^n}+\frac{\rho_{\mbox{\footnotesize    r}}}{a^{n+1}},
\ee
where $\rho_{\mbox{\footnotesize    m}}>0,\rho_{\mbox{\footnotesize    r}}>0$ are constants, in (\ref{5}). This corresponds to the mixture of pressure-free matter, $\rho={\rho_{\mbox{\footnotesize    m}}}/{a^n}$,
and radiation matter, $\rho={\rho_{\mbox{\footnotesize    r}}}/{a^{n+1}}$, valid in the universe before decoupling and
after reheating.
Hence we have
\be
\dot{a}^2=\frac{16\pi G_n}{n(n-1)}\left( \frac{\rho_{\mbox{\footnotesize    m}}}{a^{n-2}}+\frac{\rho_{\mbox{\footnotesize    r}}}{a^{n-1}}\right),
\ee
which leads us to the integral
\be 
\int a^{\frac{1-n}2}\sqrt{\rho_{\mbox{\footnotesize m}} a+\rho_{\mbox{\footnotesize r}}}\,\dd a.
\ee
 It is clear that the Chebyshev criterion is
satisfied for any integer $n\geq1$. In other words, the model can be explicitly integrated in any dimensions.

 For example, when $n=3$,  we readily see that the solution with the initial condition $a(0)=0$ is given by
\be
(\rho_{\mbox{\footnotesize    m}} a+\rho_{\mbox{\footnotesize    r}})^{\frac32}-3\rho_{\mbox{\footnotesize r}} (\rho_{\mbox{\footnotesize    m}} a+\rho_{\mbox{\footnotesize    r}})^{\frac12}+2\rho_{\mbox{\footnotesize    r}}^{\frac32}
=\sqrt{6\pi G_3}\rho_{\mbox{\footnotesize    m}}^2 t,\quad t\geq0.
\ee

\section{Integration in conformal time}
\setcounter{equation}{0}

It is also interesting to investigate the issue of integrating the Friedmann equations in terms of  conformal time $\eta$ which is
related to the cosmological time $t$ by the scale factor:
 $ \dd t=a\dd\eta$,  since many papers employ conformal time for which
Chebeyshev's theorem also applies. In this situation, it is convenient to use $a'$ to denote $\frac{\small{\dd} a}{\small{\dd}\eta}$. Then
$a'=a\dot{a}$ and (\ref{14}) becomes
\be \label{48}
({a}')^2=\frac{16\pi G_n\rho_0}{n(n-1)}a^{-n(1+w)+4}+\frac{2\Lambda}{n(n-1)} a^4-ka^2.
\ee
In order to apply  Chebyshev's  theorem, we assume that 
$w$ is a rational number and consider $k=0$ and
$\Lambda=0$ separately. When $k=0$, the conformal time version of (\ref{15}) reads
\be \label{49}
a'=\pm a^2\sqrt{c_0 a^{-n(1+w)}+\Lambda_0 },
\ee
whose integration is
\be \label{50}
\pm\int a^{-2}\left(c_0 a^{-n(1+w)}+\Lambda_0 \right)^{-\frac12}\dd a=\eta+C.
\ee
Consequently  Chebyshev's  theorem indicates that, when $\Lambda\neq0$, the left-hand side of (\ref{50}) is elementary if
and only if $\frac1{n(1+w)}$ or $\frac1{n(1+w)}-\frac12$ is an integer (again we exclude the trivial case $w=-1$), or more explicitly, $w$ satisfies one of the following
conditions:
\bea
w&=&-1+\frac1{nN},\quad N=\pm1,\pm2,\dots;\label{x4}\\
w&=&-1+\frac1{n\left(N+\frac12\right)},\quad N=0,\pm1,\pm2,\dots.
\eea

As a simple illustration, we choose $N=1$ in (\ref{x4}) so that $w=\frac1n-1$. Then (\ref{49}) becomes
\be
a'=\pm a^2\sqrt{c_0 a^{-1}+\Lambda_0},
\ee
which can be integrated to yield the solution
\be
\frac{c_0}a+\Lambda_0=\frac{c_0^2}4(\eta+C)^2,
\ee
where $C$ is an integrating constant.

On the other hand, however, when $\Lambda=0$, the Friedmann equation is
\be \label{53}
a'=\pm a\sqrt{c_0 a^{-n(1+w)+2}-k},
\ee
which in view of  Chebyshev's  theorem can be
integrated in terms of elementary functions when $w$ is
any rational number. In fact, as before, (\ref{53}) may actually be integrated to yield its exact
solution expressed in elementary functions for any $w$:
\be \label{57}
\pm\left(1-\frac12 n(1+w)\right)\eta+C=\left\{\begin{array}{rll} &-\frac1v,&\quad k=0;\\
&&\\
&\arctan v,&\quad k=1;\\
&&\\
&\frac12\ln\left|\frac{v-1}{v+1}\right|,&\quad k=-1,\end{array}\right.
\ee
where $C$ is an integrating constant and
\be
v=\sqrt{c_0 a^{-n(1+w)+2}-k}\quad \mbox{ or }\quad a^{-n(1+w)+2}=\frac1{c_0} (v^2+k).
\ee

To obtain inflationary solutions, we assume
\be
n(1+w)>2,
\ee
which contains the dust and radiation matter situations since $n\geq3$. We assume the initial
condition $a(0)=0$. When $k=0$, (\ref{57}) gives us the solution
\be\label{xx1}
a^{n(1+w)-2}(\eta)=\frac{4\pi G_n\rho_0}{n(n-1)}\left(n(1+w)-2\right)^2\eta^2,\quad \eta\geq0.
\ee
When $k=1$, (\ref{57}) renders the solution
\be
a^{n(1+w)-2}(\eta)=\frac{16\pi G_n \rho_0}{n(n-1)}\sin^2\left(\frac12(n[1+w]-2)\eta\right),\quad  \eta\geq0,
\ee
which gives rise to a periodic universe. When $k=-1$, (\ref{57}) yields the solution
\be
a^{n(1+w)-2}(\eta)=\frac{16\pi G_n \rho_0}{n(n-1)}\sinh^2\left(\frac12(n[1+w]-2)\eta\right),\quad  \eta\geq0,
\ee
which leads to an inflationary universe. It is interesting to note that the closed universe
situation here ($k=1,\Lambda=0$), in conformal time, is comparable to the flat universe with a negative cosmological
constant ($k=0,\Lambda<0$), in cosmological time, and the open universe situation here ($k=-1,\Lambda=0$),
in conformal time, is comparable to the flat universe with a positive cosmological constant ($k=0,
\Lambda>0$),
also in cosmological time.

It is immediate to check that, when $k=0$ and $\Lambda=0$, the solutions (\ref{x1}) and (\ref{xx1})
are the same since the cosmological time $t$ and conformal time $\eta$ are related through
the equation
\be
t^{n(1+w)-2}=4\pi G_n\rho_0(1+w)^2 \left(\frac n{n-1}\right)\left(1-\frac2{n(1+w)}\right)^{n(1+w)}
\eta^{n(1+w)}.
\ee
\medskip

Thus, in particular, we see that,
regarding integrability of the Friedmann equation in terms of elementary functions, the conclusions in the cosmological and conformal time situations in view of  Chebyshev's  theorem are elegantly
complementary.

\medskip

The research of Chen was supported in part by
Henan Basic Science and Frontier Technology Program
Funds under Grant No. 142300410110.  Yang was partially supported by National Natural Science Foundation of China under Grant No. 11471100. We are grateful to
John Barrow for  helpful conversations and for bringing \cite{Harrison}
to our attention.
\small{

}


\begin{thebibliography}{99}


\bibitem{Barrow}
 Barrow J D, {\em Relativistic cosmology and the regularization of orbits}, 1993
{\em The Observatory} {\bf113} 210.

\bibitem{Gibbons}
Gibbons G W, {\em Roulettes, Fermat's principle, soap films and Friedmann--Lemaitre cosmologies}, 2012  unpublished notes.

\bibitem{SK}
Stephani H,  Kramer D,  Maccallum M,  Hoenselaers C, and  Herlt E, {\em Exact Solutions of
Einstein's Field Equations}, 2003 Cambridge U. Press, Cambridge, U. K.


\bibitem{F}
 Faraoni V, {\em Solving for the dynamics of the universe},  1999 {\em Amer. J. Phys.} {\bf67} 732.

\bibitem{L}
 Lima J A S, {\em Note on solving for the dynamics of the universe}, 2001 {\em Amer. J. Phys.} {\bf69}
 1245.

\bibitem{UL}
 Uzan J-P and  Lehoucq R, {\em A dynamical study of the Friedmann equations}, 2001
{\em Eur. J.  Phys.} {\bf22}
 371.

\bibitem{Linder}
 Linder E V,
{\em Exploring the expansion history of the universe}, 2003
{\em Phys. Rev. Lett.} {\bf90}  091301.

\bibitem{CDS}
 Caldwell R R,  Dave R, and  Steinhardt P J,
{\em Cosmological imprint of an energy component with general equation of state}, 1998
{\em Phys. Rev. Lett.} {\bf80} 1582.

\bibitem{van}
 van Holten J W,
{\em Cosmological Higgs fields},
2002 {\em Phys. Rev. Lett.} {\bf89} 201301.



\bibitem{W}
 Wald R M, {\em General Relativity}, 1984 U. Chicago Press, Chicago and London.

\bibitem{Harrison}  Harrison E R, {\em Classification of uniform cosmological models}, 1967
 {\em Mon. Not. R. Astr. Soc.} {\bf 137}   69.

\bibitem{ST}
 Sonego S and Talamini V, {\em Qualitative study of perfect-fluid Friedmann--Lemaitre--Robertson--Walker
models with a cosmological constant},  2012 {\em Amer. J. Phys.} {\bf80}  670.

\bibitem{WKS}
 Watanabe M,  Kanno S, and  Soda J,
{\em Inflationary universe with anisotropic hair},
2009 {\em Phys. Rev. Lett.} {\bf102} 191302.



\bibitem{FS0}
 Farajollahi H and  Salehi A,
{\em New approach in stability analysis: case study: Tachyon cosmology with nonminimally coupled
scalar-field matter},
2011 {\em Phys. Rev.} D {\bf83} 124042.

\bibitem{CS}
 Campos A and Sopuerta C F,
{\em Evolution of cosmological models in the brane-world scenario},
2001 {\em Phys. Rev.} D {\bf63} 104012.

\bibitem{sc1}
 Steinhardt P J  and  Turok N, {\em A cyclic model of the universe}, 2002
{\em Science}
{\bf296} 1436.



\bibitem{NOO}
Nojiri S, Odintsov S D, and Ogushi S, {\em Friedmann--Robertson--Walker brane cosmological equations
from the five-dimensional bulk (A)dS black hole},  2002 {\em Int. J. Mod. Phys. A} {\bf17}  4809.

\bibitem{FS}
Fischler W and Susskind L, {\em Holography and cosmology}, 1998 [hep-th/9806039].

\bibitem{KL}
 Kaloper N and  Linde A, {\em Cosmology versus holography}, 1999 {\em Phys. Rev.} D {\bf60}  103509.

 \bibitem{C0}
Tchebichef M P, {\em L'int\'{e}gration des diff\'{e}rentielles irrationnelles},  1853 {\em J. Math. Pures Appl.} {\bf 18} 87.

\bibitem{C1}
 Marchisotto E A and  Zakeri G-A, {\em An invitation to integration in finite terms},
1994 {\em College Math. J.} {\bf25}  295.

\bibitem{planetmath}
http://planetmath.org/integrationofdifferentialbinomial

\bibitem{Planck}
 Planck Collaboration,
{\em Planck 2013 results. I. Overview of products and scientific results}, 2013 arXiv: 1303.5062.

\bibitem{P2}
Burgess C P,  Cicoli M, and Quevedo F,
{\em String inflation after Planck 2013},
{\em J. Cosmology Astroparticle Phys.}  2013
{\bf11} 003.

\bibitem{P3}
Peebles, P J E, {\em Dark matter}, 2014 {\em Proc. Nat. Acad. Sci. U. S. A.}, to appear.



\bibitem{nature}
 de Bernardis P {\em et al}, {\em A flat Universe from high-resolution maps of the cosmic microwave background radiation},
2000 {\em Nature} {\bf404} 955.

\bibitem{science}
 Bahcall N A,  Ostriker J P,  Perlmutter S,  and Steinhardt P J,
{\em The cosmic triangle: revealing the state of the universe}, 1999
{\em Science}
{\bf284} 1481.



\bibitem{Copeland}
 Copeland E J, Sami M, and  Tsujikawa S, {\em Dynamics of dark energy},  2006 {\em Int. J. Mod. Phys.}
 D {\bf15} 1753.


 \bibitem{AA}
 Chernin A D, Nagirner D I, and Starikova S V, {\em Growth rate of cosmological perturbations in standard model: Explicit analytic solution},
 2003 {\em Astron. Astrophys.} {\bf399} 19.




\bibitem{G2}
Dunajski M and  Gibbons G W, {\em Cosmic jerk, snap and beyond}, 2008 {\em Class. Quant.Grav.}
{\bf25}  235012.

\end{thebibliography}
\end{document}